\documentclass[a4paper,UKenglish,cleveref, autoref, thm-restate]{lipics-v2021}

\title{Use and abuse of instance parameters in the Lean mathematical library}
\author{Anne Baanen}
{Department of Computer Science, Vrije Universiteit Amsterdam, The Netherlands \and \url{https://cs.vu.nl/~tbn305}}
{t.baanen@vu.nl}
{https://orcid.org/0000-0001-8497-3683}
{}
\authorrunning{T. Baanen}

\Copyright{Anne Baanen}



\category{} %optional, e.g. invited paper

\supplement{Unabridged, interactive listings are available at \url{https://github.com/lean-forward/mathlib-classes}.}
\funding{NWO Vidi grant No.\ 016.Vidi.189.037, Lean Forward}

\acknowledgements{I want to thank the anonymous reviewers and Jeremy Avigad, Jasmin Blanchette, Bryan Gin-Ge Chen, Johan Commelin, Sander Dahmen, Manuel Eberl, Rob Lewis, Assia Mahboubi, Filippo A. E. Nuccio, Kazuhiko Sakaguchi, Enrico Tassi and Eric Wieser for their helpful comments on earlier versions on this manuscript.}%optional

\nolinenumbers %uncomment to disable line numbering

\EventEditors{June Andronick and Leonardo de Moura}
\EventNoEds{2}
\EventLongTitle{13th International Conference on Interactive Theorem Proving (ITP 2022)}
\EventShortTitle{ITP 2022}
\EventAcronym{ITP}
\EventYear{2022}
\EventDate{August 7--10, 2022}
\EventLocation{Haifa, Israel}
\EventLogo{}
\SeriesVolume{237}
\ArticleNo{7}
\usepackage{amsmath}
\usepackage{amssymb}
\usepackage{hyperref}
\usepackage{xcolor}

\usepackage{xspace}
\usepackage{listings}

\newcommand{\mathlib}{\textsf{mathlib}\xspace}
\newcommand{\N}{\mathbb{N}}
\newcommand{\R}{\mathbb{R}}

\newcommand{\Z}{\mathbb{Z}}

\definecolor{keywordcolor}{rgb}{0.7, 0.1, 0.1}   % red
\definecolor{commentcolor}{rgb}{0.4, 0.4, 0.4}   % grey
\definecolor{symbolcolor}{rgb}{0.4, 0.4, 0.4}    % grey
\definecolor{sortcolor}{rgb}{0.1, 0.5, 0.1}      % green

\DeclareUnicodeCharacter{03A0}{\ensuremath{\Pi}}
\DeclareUnicodeCharacter{03B1}{\ensuremath{\alpha}}
\DeclareUnicodeCharacter{03B2}{\ensuremath{\beta}}
\DeclareUnicodeCharacter{03BB}{\ensuremath{\lambda}}
\DeclareUnicodeCharacter{03C3}{\ensuremath{\sigma}}
\DeclareUnicodeCharacter{2081}{\ensuremath{_1}}
\DeclareUnicodeCharacter{2082}{\ensuremath{_2}}
\DeclareUnicodeCharacter{2090}{\ensuremath{_a}}
\DeclareUnicodeCharacter{2097}{\ensuremath{_l}}
\DeclareUnicodeCharacter{2115}{\ensuremath{\N}}
\DeclareUnicodeCharacter{211A}{\ensuremath{\Q}}
\DeclareUnicodeCharacter{211D}{\ensuremath{\R}}
\DeclareUnicodeCharacter{2124}{\ensuremath{\Z}}
\DeclareUnicodeCharacter{220F}{\hspace{-1ex}\ensuremath{\prod}} % hack to fix listings' behaviour involving parentheses
\DeclareUnicodeCharacter{2211}{\ensuremath{\sum}}
\DeclareUnicodeCharacter{2223}{\ensuremath{\mid}}
\DeclareUnicodeCharacter{2264}{\ensuremath{\l}}
\DeclareUnicodeCharacter{2286}{\ensuremath{\subseteq}}
\DeclareUnicodeCharacter{22A4}{\ensuremath{\top}}
\DeclareUnicodeCharacter{22A5}{\ensuremath{\bot}}
\DeclareUnicodeCharacter{22EF}{\ensuremath{\cdots}}

\begin{document}

\maketitle

\begin{abstract}
The Lean mathematical library \mathlib features extensive use of the typeclass pattern for organising mathematical structures,
based on Lean's mechanism of instance parameters.
Related mechanisms for typeclasses are available in other provers including Agda, Coq and Isabelle with varying degrees of adoption.
This paper analyses representative examples of design patterns involving instance parameters in the current Lean 3 version of \mathlib,
focussing on complications arising at scale and how the \mathlib community deals with them.
\end{abstract}

\section{Introduction}

An essential part of a mathematical library is a discipline for representing structures,
such as an algebraic hierarchy including monoids, groups, rings and fields,
or a hierarchy of spaces including topological spaces and metric spaces.
Many design patterns have been proposed to enable the theory of a general structure (such as monoids) to be applied seamlessly to more specific structures (such as fields),
including canonical structures in Coq~\cite{Coq-Packed-classes}, locales in Isabelle~\cite{Isabelle-locale, ballarin-locales}, unification hints in Matita~\cite{matita-unification-hints} and attributed types in Mizar~\cite{Mizar-attributed-type}.

The Lean mathematical library \mathlib~\cite{mathlib} has settled on the use of the \emph{typeclass}~\cite{typeclasses-haskell} pattern for representing structures,
implemented through Lean's mechanism of \emph{instance parameters}~\cite{lean-prover, lean-elaboration}.
Typeclasses were invented by Wadler to provide ad hoc polymorphism in Haskell~\cite{typeclasses-haskell}.
Similar mechanisms can now be found in programming languages including Idris~\cite{Idris}, Rust~\cite[Chapter 6.11]{Rust-Reference} and Scala~\cite{Scala-Implicits},
and interactive theorem provers including Agda~\cite{Agda-Instance-Arguments}, Coq~\cite{Coq-first-class-type-class} and Isabelle~\cite{Isabelle-typeclasses}.
Class-based libraries of comparable complexity to \mathlib have previously been developed by Hölzl, Immler and Huffman~\cite{Isabelle-HOL-analysis-classes} for analysis in Isabelle/HOL
and by Spitters and Van der Weegen~\cite{Coq-typeclass-hierarchy} for an algebraic hierarchy in Coq.
As of February 2022, \mathlib contains over 600 classes and over 8000 instances.

In the various languages implementing a mechanism analogous to typeclasses, there is also a variety of syntax and semantics for the parts of this mechanism.
In this paper, I will try to avoid confusion by using the terminology ``instance parameters'' when emphasising Lean 3's specific implementation,
while ``typeclass'' refers to a design pattern that is implemented in Lean through instance parameters.
This is analogous to the distinction drawn in the Scala documentation between its mechanism of ``implicit parameters'' and the typeclass pattern built with that mechanism.

This paper combines my original research with a survey of the \mathlib community's experience in developing a class-based hierarchy,
emphasising design patterns, issues arising from the use of classes and the trade-offs available for resolving these issues.
Researching how to develop and maintain a large library of mathematical structures
has led me to develop a number of typeclass-based patterns in Lean that have been added to \mathlib.
Given that \mathlib is expected to upgrade soon from Lean 3 to Lean 4~\cite{lean-4}, an upgrade that promises changes in Lean's support for typeclasses~\cite{lean-tabled-typeclasses},
now seems a good time to discuss what is achievable with the current mechanism.
I have organized this paper around a representative example for each topic, based on \mathlib source code.
Unabridged, interactive versions of the listings are available at \url{https://github.com/lean-forward/mathlib-classes}.

\section{Basic instance parameters in Lean 3} \label{sec:basics}
Lean provides the typeclass pattern through \emph{instance parameters}, a mechanism and implementation closely resembling the same facilities in Coq~\cite{Coq-first-class-type-class}.
Like Coq, Lean is a dependently typed language based on the calculus of constructions.
Lean has a hierarchy of universes \lstinline{Sort 0 : Sort 1 : Sort 2 : ...}, where \lstinline{Sort 0} is more often written as \lstinline{Prop} and
\mbox{\lstinline{Sort (u + 1)}} is written \lstinline{Type u} or, leaving \lstinline{u} implicit, \lstinline{Type*}.
The bottom universe \lstinline{Prop} is an impredicative type of propositions that has definitional proof irrelevance.
% The notation \lstinline{Type*} stands for ``\lstinline{Type u} for an arbitrary universe level \lstinline{u}''; leaving the level implicit in this way typically suffices.

Let us start with the following two Lean declarations, not found in \mathlib, showing the main forms that parameters can take in Lean.
\begin{lstlisting}
def sub {A : Type*} [add_group A] (a b : A) : A :=
add a (neg b)

lemma sub_eq_add_neg {A : Type*} [add_group A] (a b : A) :
  sub a b = add a (neg b) := by refl
\end{lstlisting}
The round brackets mark explicit parameters to be supplied when applying the lemma,
the curly brackets mark implicit parameters inferred through unification,
while square brackets mark the instance parameters (for which supplying a name is optional);
these are used here to specify a typeclass constraint on the type \lstinline{A}.
Thus the term \lstinline{sub a b} specifies only the \lstinline{(a b : A)} parameters to \lstinline{sub},
leaving the remaining parameters to be supplied by the elaborator.
These parameters are then passed on to the calls to \lstinline{neg} and \lstinline{add} in the body of \lstinline{sub}.
There is no relevant distinction between the keywords \lstinline{def} and \lstinline{lemma} for our purposes,
apart from indicating whether the declaration exists in a \lstinline{Type} or the \lstinline{Prop} universe.

The elaborator supplies values to instance parameters through \emph{instance synthesis}:
the parameters to the current declaration and all declarations in the global context which have been marked with the keyword \lstinline{instance} are considered in turn as candidates.
Each candidate instance is type-checked against the goal, and the first candidate where the types unify is returned.
For example, defining \lstinline{add_group ℤ} instance allows us to subtract two integers using the \lstinline{sub} operator we defined above:
\begin{lstlisting}
instance : add_group ℤ := sorry -- implementation omitted
lemma subtraction_example : (sub 42 37 : ℤ) = 5 := by refl
\end{lstlisting}
% Types of instance parameters, such as \lstinline{add_group} above, are required to have the \lstinline{@[class]} annotation.

Instance parameters are considered a form of implicit parameters, and can thus be made explicit using the \lstinline{@} operator:
\begin{lstlisting}
#check sub -- sub : ?M_1 → ?M_1 → ?M_1
#check @sub -- sub : Π {A : Type*} [add_group A], A → A → A
\end{lstlisting}
Here \lstinline{?M_1} stands for a free metavariable that the elaborator could not (yet) fill in.

Instance declarations can themselves have their own instance parameters.
For example, the subsets of a monoid form a monoid under pointwise operations, which we can express as
\begin{lstlisting}
instance pointwise_monoid {A : Type*} [monoid A] : monoid (set A) :=
{ mul := λ X Y, { (x * y) | (x ∈ X) (y ∈ Y) },
  mul_assoc := λ _ _ _, set.image2_assoc mul_assoc,
  ..sorry /- further fields omitted -/ }
\end{lstlisting}
When the synthesis of an instance of \lstinline{monoid (set A)} tries to apply the \lstinline{pointwise_monoid} instance,
the elaborator will recurse and try to synthesize the \lstinline{monoid A} dependency;
if this instance is not found, search backtracks and continues with the next \lstinline{monoid (set A)} instance;
if the entire search tree is exhausted, synthesis has failed.
Since instances' types are unified with the goal, we can view the synthesis algorithm as performing recursion on the term structure of the goal.

\subsection{Class definitions}
The classes themselves are expressed as records, i.e. dependent tuples, with the class fields as projections taking instance parameters.
Classes use the same syntax as record types in Lean, only differing in using the keyword \lstinline{class} instead of \lstinline{structure}.
Dependent types mean data- and proof-carrying fields are expressed in the same way;
Section~\ref{sec:diamond} discusses some usage differences between data and proofs.
Thus, a definition for \lstinline{add_group} could look like:
\begin{lstlisting}
class add_group (A : Type*) : Type* :=
(zero : A) (neg : A → A) (add : A → A → A)
(add_assoc : ∀ (x y z : A), add x (add y z) = add (add x y) z)
(zero_add : ∀ (x : A), add zero x = x)
(neg_add : ∀ (x : A), add (neg x) x = zero)
\end{lstlisting}
%Since there are fields of type \lstinline{A : Type*}, the record type \mbox{\lstinline{add_group A}} will be of type \lstinline{Type*} too.
%Lean also allows classes to inhabit the universe \lstinline{Prop} of proof-irrelevant propositions.
%The \lstinline{class} keyword automatically uses instance parameters in the projections, in effect generating the following code:
The projections of a class automatically use instance parameters, generating the declarations:
\begin{lstlisting}
def add_group.zero {A : Type*} [h : add_group A] : A := h.1
def add_group.neg {A : Type*} [h : add_group A] : A → A := h.2
def add_group.add {A : Type*} [h : add_group A] : A → A → A := h.3

def add_group.add_assoc {A : Type*} [h : add_group A] :
  ∀ (x y z : A), add_group.add x (add_group.add y z) =
    add_group.add (add_group.add x y) z) := h.4
-- and so on.
\end{lstlisting}
The instance synthesis algorithm also allows instances for non-record types.
In practice most classes in \mathlib are record types and indeed Lean 4 will make the use of record types for classes obligatory to simplify the elaborator.

\subsection{Subclassing}
The mechanisms described above result in two patterns for subclass definitions, with important distinctions in semantics.
\emph{Unbundled subclasses} take superclasses as instance parameters to the class declaration.
To define abelian groups as a subclass of additive groups, we write
\begin{lstlisting}
class add_comm_group (A : Type*) [add_group A] : Type* :=
(add_comm : ∀ (x y : A), add x y = add y x)
\end{lstlisting}
Elaboration of the expression \lstinline{add_comm_group A}, e.g.~in a parameter \mbox{\lstinline{[add_comm_group A]}},
requires the synthesis of an \lstinline{add_group A} instance occurring as parameter to \lstinline{add_comm_group}.
We make this instance available by adding it as another instance parameter,
so all results on abelian groups take the two instance parameters \lstinline{[add_group A] [add_comm_group A]};
long inheritance chains cause long parameter lists.
Similarly, declaring an \lstinline{add_comm_group A} instance requires a previous declaration of an \lstinline{add_group A} instance.

In contrast, \emph{bundled subclasses} make use of instance synthesis to access the superclass, only requiring an instance parameter for the subclass.
A bundled subclass contains an instance of its superclass as a record field.
Lean's \lstinline{extends} keyword provides syntax sugar for the construction:
\begin{lstlisting}
class add_comm_group (A : Type*) extends add_group A :=
(add_comm : ∀ (x y : A), add x y = add y x)
\end{lstlisting}
This has analogous effects to writing
\begin{lstlisting}
class add_comm_group (A : Type*) : Type* :=
(to_add_group : add_group A)
(add_comm : ∀ (x y : A),
  @add_group.add A to_add_group x y = @add_group.add A to_add_group y x)
-- Register the projection as an instance:
attribute [instance] add_comm_group.to_add_group
\end{lstlisting}
When we look at the synthesis of an \lstinline{add_group A} instance, we can see the instance \lstinline{add_comm_group.to_add_group} triggers a recursive search for an \lstinline{add_comm_group A} instance;
in this way subclass instances automatically provide access to declarations on the superclass.
Both subclass patterns are used in \mathlib; the following sections discuss reasons for choosing one over the other in a given situation.

\subsection{Extensions of the typeclass pattern}

Beyond expressing the basic typeclass patterns above, Lean's instance parameters provide a considerable amount of flexibility.
Classes do not have to be parametrized over exactly one type, unlike what the phrase ``typeclass'' suggests,
and the only restriction on \lstinline{instance} declarations is that the head symbol of its return type is declared to be a class.
In Haskell terminology, all the following extensions are allowed: constrained class method types, flexible contexts, flexible instances, incoherent instances, multi-parameter classes (including nullary classes), overlapping instances, quantified constraints, type synonym instances, undecidable instances.
Most of these extensions find uses throughout \mathlib.

A notable difference compared with Agda is that Lean allows overlapping instances, thus enabling the diamond inheritance pattern of Section~\ref{sec:monoid}.
Compared with Isabelle, Lean permits multi-parameter classes, as we will see in Section~\ref{sec:multiparam}; on the other hand Isabelle provides multi-parameter hierarchies through locales~\cite{ballarin-locales}.
Compared with Coq, Lean adds a limited syntax for expressing functional dependencies, also discussed in Section~\ref{sec:module-and-vector-space}.

Since Lean is a dependently typed language, parameters to classes are not restricted to types.
For example, \mathlib uses a typeclass parametrized over a natural number $p$ to express the characteristic of a ring:
\begin{lstlisting}
class char_p (R : Type*) [semiring R] (p : ℕ) : Prop :=
(cast_eq_zero_iff : ∀ (x : ℕ), (coe x : R) = 0 ↔ p ∣ x)
\end{lstlisting}
% Actually [add_monoid R] [has_one R], but let's not get into that too closely.
The \lstinline{char_p} class is an example of \emph{a proof mixin}.
Section~\ref{sec:mixin} discusses this pattern further.

Since type-checking is used to match candidate instances with a synthesis goal,
the synthesis algorithm works up to definitional equality.
For example, since \lstinline{2+2 : ℕ} is definitionally equal to \lstinline{4},
Lean finds the instance \lstinline{zmod.char_p 4 : char_p (zmod 4) 4} for the goal \lstinline{char_p (zmod 4) (2 + 2)}.
Thus, typeclasses in Lean are coupled to a built-in syntactic notion of equality.

This combination of features means that instance parameters can be used for small-scale automation,
since the instance synthesis mechanism provides a search tactic for definite Horn clauses:
a clause of the form $C = (\bigwedge_{i} P_i(\vec{t})) \to Q(\vec{t})$, where the $P_i$ and $Q$ are (not necessarily distinct) predicates and $\vec{t}$ a vector of terms,
translates to an instance of the form \lstinline{instance C [P_1 t_1 ... t_k] ... [P_n t_1 ... t_k] : Q t_1 ... t_k}.
The branching depth-first nature of the synthesis algorithm has to be kept in mind during design in order to assure acceptable performance, as we will see in Section~\ref{sec:performance}.

\section{\lstinline{has_mul}: notation typeclass} \label{sec:notation}

The typeclass pattern is used throughout \mathlib for operator overloading,
in much the same role that classes were originally introduced in Haskell.
Generally, such a notation typeclass has one type parameter \lstinline{α : Type*}
and contains fields which carry only data.
A basic example is the definition of the multiplication operator \lstinline{*} in core Lean:
\begin{lstlisting}
class has_mul (α : Type*) := (mul : α → α → α)
infix * := has_mul.mul
\end{lstlisting}
The \lstinline{infix} command adds the notation \mbox{\lstinline{a * b}} for \lstinline{has_mul.mul a b}.
% Analogous instances are provided in the core Lean libraries or \mathlib for algebraic operations such as addition $+$, multiplicative inverses ${}^{-1}, an additive identity $0$ and a multiplicative identity $1$.

These notations are not directly coupled to the algebraic hierarchy:
the \lstinline{has_inv} class providing ${}^{-1}$ notation for the multiplicative inverse does not have any fields requiring a multiplicative group structure.
However, in practice such notations are often provided through inheritance from an instance of a proof-carrying class in the algebraic hierarchy.

Lean uses classes to implement implicit coercions in the style of Saïbi~\cite{Coq-Typing-With-Inheritance}.
Whenever the elaborator encounters a term \lstinline{t : A} that is instead expected to have type \lstinline{B},
it replaces \lstinline{t} with \lstinline{@coe A B _ t},
where the \lstinline{_} marks an instance parameter of type \lstinline{has_lift_t A B}.
Similarly, when a term \lstinline{f : F} produces a type error because it is expected to have a dependent function type,
it is replaced with \lstinline{coe_fn f} (where \mbox{\lstinline!coe_fn \{F A\} [has_coe_to_fun F A]!} has type \lstinline!Π (f : F), A f!),
and when \lstinline{t} is expected to be of the form \lstinline{Sort u} (that is, either \lstinline{Type v} if \lstinline{u = v+1}, or \mbox{\lstinline{Prop} if \lstinline{u = 0})}, it is replaced with \lstinline{coe_sort t}
(where \mbox{\lstinline!coe_sort \{A\} [has_coe_to_sort A] :!} \mbox{\lstinline!Sort u!}).
Such coercions are essential for \mathlib's design of morphisms and subobjects, as we will see in Section~\ref{sec:hom-subobject}.

\section{\lstinline{comm_monoid}: algebraic hierarchy class} \label{sec:monoid} \label{sec:alg-hierarchy}

The algebraic hierarchy in \mathlib is built using typeclasses,
based on the notation typeclasses discussed in the previous section.
Similar class-based hierarchies exist in \mathlib for topics including orders, topology and analysis,
and all the hierarchies are connected throughout.
%For ease of presentation, this section will focus on the algebraic part of the hierarchy.
As an example, the \lstinline{comm_monoid} typeclass is implemented in \mathlib essentially as follows:
\begin{lstlisting}
set_option old_structure_cmd true -- explained below

class semigroup (G : Type*) extends has_mul G :=
(mul_assoc : ∀ a b c : G, a * b * c = a * (b * c))

class mul_one_class (M : Type*) extends has_one M, has_mul M :=
(one_mul : ∀ a : M, 1 * a = a) (mul_one : ∀ a : M, a * 1 = a)

class comm_semigroup (G : Type*) extends semigroup G :=
(mul_comm : ∀ a b : G, a * b = b * a)

class monoid (M : Type*) extends semigroup M, mul_one_class M

class comm_monoid (M : Type*) extends monoid M, comm_semigroup M
\end{lstlisting}
While \lstinline{comm_monoid} is considered to sit low in the \mathlib algebraic hierarchy,
its definition already depends on seven ancestor classes in a complicated diamond inheritance pattern.
Multiple inheritance paths result in two instances of \lstinline{has_mul} for each \lstinline{monoid} instance, thus requiring support for overlapping instances.
We can also see that \mathlib prefers bundled inheritance in the algebraic hierarchy, incorporating ancestor classes' fields rather than taking superclasses as instance parameters.
This choice is further explained in Section~\ref{sec:performance}.

The various hierarchies in \mathlib are interwoven through multiple inheritance.
Thus the hierarchy of order structures such as partial orders, linear orders and lattices (extending the notation typeclasses \lstinline{has_le} providing \lstinline{(≤)} and \lstinline{has_lt} providing \lstinline{(<)}),
is combined with the algebraic hierarchy into a hierarchy of ordered algebraic structures from partially ordered commutative monoids up to linearly ordered fields.

The first line \lstinline{set_option old_structure_cmd true} switches between two representations
of ancestors for \lstinline{structure} and \lstinline{class} declarations:
under the default, ``new'' structure behaviour,
\lstinline{monoid M} would contain two fields, of type \lstinline{semigroup M} and \lstinline{mul_one_class M}, each of which carries its own distinct \lstinline{has_mul} field.
Thus the \lstinline{mul_assoc} field inherited from \lstinline{semigroup} would refer to a multiplication operation
other than the multiplication of \lstinline{one_mul} %and \lstinline{mul_one}
inherited from \lstinline{mul_one_class};
the resulting class with two binary operators would not actually specify monoids.
Indeed, Lean will detect such ambiguities and produce an error if a ``new'' structure inherits conflicting field names.

The ``old'' structure behaviour avoids this issue by copying all fields from the ancestor structure into the child structure, skipping duplicates,
so that \lstinline{monoid} only has one \lstinline{mul} field.
Compare the following two desugarings of \lstinline{extends}:
\begin{lstlisting}
class monoid_new (M : Type*) :=
(to_semigroup : semigroup M)
(to_mul_one_class : mul_one_class M)

class monoid_old (M : Type*) :=
(mul : M → M → M) (mul_assoc : ∀ a b c : M, a * b * c = a * (b * c))
(one : M) (one_mul : ∀ a : M, 1 * a = a) (mul_one : ∀ a : M, a * 1 = a)
\end{lstlisting}

Lean 4 only implements the ``new'' structure command since it directly allows projecting to ancestor structures, %\footnote{\url{https://github.com/leanprover/lean/wiki/Refactoring-structures\#encoding-inheritance}},
adding support for diamond inheritance through automatically inheriting from the common ancestor and copying the remaining fields.

In the terminology of Coq's Hierarchy Builder~\cite{hierarchy-builder}, the typeclasses are specified in terms of \emph{mixins}:
the packages of operations and properties available for a given structure.
Like Hierarchy Builder provides for mixins, projections from a subclass to its immediate superclasses are automatically generated as instances.
There is no explicit concept in Lean corresponding to Hierarchy Builder's \emph{factories} or \emph{builders}.
To manually construct a subclass instance given a superclass or project a subclass into a superclass,
users can apply the notation \lstinline!.. s!, which extends a constructor's argument list by copying the relevant fields out of a tuple \lstinline{s}.

In general, \mathlib's hierarchy is extended when the mathematics requires it,
so there are many parts of the hierarchy that do not form a boolean algebra.
Thus there is no \lstinline{comm_mul_one_class} forming the direct subclass of \lstinline{comm_semigroup} and \mbox{\lstinline{mul_one_class},}
nor is there a \lstinline{comm_mul_class} that provides the \lstinline{mul_comm} field by itself.
Adding an intermediate class to the hierarchy is a straightforward process of moving over the fields and modifying the \lstinline{extends} clauses,
as recently happened with the addition of \lstinline{mul_one_class}.%
%\footnote{\url{https://github.com/leanprover-community/mathlib/pull/6865}}

The relative ease of modification means the hierarchy does not need to be designed up front for all potential usages.
This stands in contrast to the situation for packed classes, where refactoring the hierarchy involves a deep understanding of the details involved or the usage of a tool such as Hierarchy Builder,
to ensure consistency such as the uniqueness of a join for any two structures applied to the same type~\cite{Coq-Validating-Mathematical-Structures}.
Careful design is still needed for instances to avoid certain cases of drastic slowdowns, as seen in Section~\ref{sec:performance}.

In addition to the above multiplicative hierarchy, \mathlib includes an isomorphic additive hierarchy differing only in notation:
the definition of \lstinline{add_monoid} renames \lstinline{(*)} and \lstinline{1} to \lstinline{(+)} and \lstinline{0}.
A metaprogram \lstinline{to_additive} creates an appropriately renamed duplicate additive version of declarations.
The duplicate notation is required for the definition of the \lstinline{semiring} class in \mathlib,
since bundled class inheritance cannot express the fact that the additive and multiplicative structures of a semiring both form a monoid.
In comparison to \mathlib's ad hoc solution, Isabelle's locales support different notations automatically, since the operations of a target context can be renamed in sublocale declarations or instantiations~\cite{ballarin-locales-algebra}.

\section{\lstinline{module}: multi-parameter classes} \label{sec:module-and-vector-space} \label{sec:multiparam}

In algebra, a (\emph{left semi}-)\emph{module} is an additive commutative monoid \lstinline{M} that is acted on by a semiring \lstinline{R} through scalar multiplication,
satisfying certain axioms;
the concept generalizes vector spaces by replacing the field of scalars by an arbitrary semiring.
Modules are available in the \mathlib algebraic hierarchy in full generality as a multi-parameter typeclass depending on both \lstinline{R} and \lstinline{M}.
Following the pattern of monoids, the base class introduces notation, and is subclassed to add the axioms of the structure:
\begin{lstlisting}
class has_scalar (α β : Type*) : Type* :=
(smul : α → β → β)
infix • := has_scalar.smul

class mul_action (M A : Type*) [monoid M] extends has_scalar M A :=
(one_smul : ∀ (x : A), (1 : M) • x = x)
(mul_smul : ∀ (r s : M) (x : A), (r * s) • x = r • (s • x))

class distrib_mul_action (M A : Type*) [monoid M] [add_monoid A]
  extends mul_action M A :=
(smul_add : ∀ (r : M) (x y : A), r • (x + y) = r • x + r • y)
(smul_zero : ∀ (r : M), r • (0 : A) = 0)
\end{lstlisting}
\pagebreak % please keep `smul_zero' on the same page
\begin{lstlisting}
class module (R M : Type*) [semiring R] [add_comm_monoid M]
  extends distrib_mul_action R M :=
(add_smul : ∀ (r s : R) (x : M), (r + s) • x = r • x + s • x)
(zero_smul : ∀ (x : M), (0 : R) • x = 0)
\end{lstlisting}
Compare this to the class-based analysis library in Isabelle/HOL, where the absence of multi-parameter classes means only real numbers appear as scalars~\cite{Isabelle-HOL-analysis-classes};
Isabelle instead provides multi-parameter locales~\cite{ballarin-locales}.

Vector spaces do not have a separate definition in \mathlib since they only replace the ring axioms on the scalars with field axioms,
while the fields of the \lstinline{module} class are unchanged.
Instead, a \lstinline{K}-vector space \lstinline{V} is denoted through parameters \mbox{\lstinline{[field K]}} \mbox{\lstinline{[add_comm_group V] [module K V]}}.
In order to make vector spaces more discoverable for users,
the \mathlib community has been discussing a system of parameter-level abbreviations,
so that \lstinline{[vector_space K V]} expands into \mbox{\lstinline{[field K]}} \mbox{\lstinline{[add_comm_group V]}} \mbox{\lstinline{[module K V]}}.

\subsection{Dangerous instances} \label{sec:dangerous-instance}
We see here that the \lstinline{module} hierarchy uses a mix of bundled and unbundled inheritance,
unlike \lstinline{comm_monoid} which solely uses bundled inheritance.
This follows the rule of bundling only if the superclass has a superset of the subclass's parameters;
otherwise the generated instance would be a \emph{dangerous instance} where some parameters are undetermined.
Namely, declaring that \lstinline{module R M extends} \lstinline{add_comm_monoid M} would generate the following instance:
\begin{lstlisting}
instance module.to_add_comm_monoid {R M : Type*} [module R M] :
  add_comm_monoid M := sorry
\end{lstlisting}
Now instance synthesis for \lstinline{add_comm_monoid M} will lead to a search for \lstinline{module ?M_1 M},
where \lstinline{?M_1} is a free metavariable.
Since unification in the elaborator can call instance synthesis, without backtracking,
finding the wrong instance for an underspecified goal may cause unification to fail where another instance would have worked.
%In general, parameters to an instance should be inferrable through either unification with the goal type, or uniquely determined through instance synthesis.

Such dangerous instances with free variables in their constraints can be remedied in various ways.
If, as above, the instance derives from a subclass constraint involving the \lstinline{extends} keyword,
the constraint is instead expressed through an instance parameter on the subclass;
this implies no dangerous instance is generated to express inheritance.
The main drawback is that this mix of unbundled and bundled inheritance is more confusing and less natural than the approach used in the MathComp library,
where canonical structures allow bundling the additive monoid structure and the $R$-module structure~\cite{MathComp}.

If the free parameter is uniquely determined by the choice of the bound parameters,
we can register this functional dependency with the \lstinline{out_param} construction.
Lean assigns all parameters as in-parameters, unless explicitly marked as \lstinline{out_param},
in contrast to the automatic determination of the direction in Coq.
For example, we can make \lstinline{R} a functional dependency of \lstinline{M} by instead defining:
\begin{lstlisting}
class module (R : out_param Type*) (M : Type*) [semiring R] := -- etc.
\end{lstlisting}
The elaborator replaces all out parameters in the synthesis goal with a free metavariable,
which is filled by unifying the goal with the type of the candidate instance.
%In other words, \lstinline{out_param} switches parameters from being supplied by the goal type to being supplied by the instance type.
For \lstinline{module} this functional dependency is not acceptable
since each \lstinline{add_comm_monoid M} has an instance \lstinline{add_comm_monoid.nat_module : module ℕ M} reflecting the natural $\N$-module structure (see also Section~\ref{sec:nsmul}),
which would be incompatible with an $R$-module structure for other semirings $R$.
Moreover, the instance \lstinline{add_comm_monoid.nat_module} provides a second reason that bundled inheritance is unsuitable for the subclass relation of \lstinline{add_comm_monoid} and \lstinline{module}:
it would form a loop with the instance \lstinline{module.to_add_comm_monoid}.

A final way to resolve dangerous instances is to remove the \lstinline{instance} keyword so that it does not participate in synthesis.
\mathlib takes this approach when stating the theorem that any module over a ring has additive inverses:
\begin{lstlisting}
def module.add_comm_monoid_to_add_comm_group (R M : Type*)
  [ring R] [add_comm_monoid M] [module R M] :
  add_comm_group M := sorry -- proof omitted
\end{lstlisting}
To provide \lstinline{add_comm_group} instances when \lstinline{R} is known,
we can still make use of the instance \lstinline{add_comm_monoid_to_add_comm_group} in a separate \lstinline{instance} declaration.

\section{\lstinline{monoid_hom_class}: generic bundled morphisms} \label{sec:hom-subobject}

The representation of morphisms such as group homomorphisms or linear maps has changed repeatedly in \mathlib,
is still not unified and is still undergoing refactors.
The main issue complicating the design is the trade-off between generality and ease of inference.
The author of this paper has designed a pattern providing bundled morphisms with some of the advantages lost during the move from unbundled morphisms,
by making theorems generic over types of morphisms.
The same pattern works for subobjects, replacing ``morphism'' with ``subobject'' and ``a map preserving an operation'' with ``a set closed under an operation''.

\subsection{Unbundled morphisms}
The original design of algebraic homomorphisms in \mathlib did not bundle maps in the same structure as their properties,
allowing any function \lstinline{f : R → S} to be used as a ring homomorphism if an \lstinline{is_monoid_hom f} instance was available.
The \lstinline{is_ring_hom} predicate stated \lstinline{f} preserves the ring operations \lstinline{*}, \lstinline{+}, \lstinline{1} and \lstinline{0}.
Instances were available for the common operations, except composition:
\begin{lstlisting}
class is_monoid_hom {M N : Type*} [monoid M] [monoid N] (f : M → N) : Prop :=
(map_mul : ∀ x y : M, f (x * y) = f x * f y)
(map_one : f 1 = 1)

class is_ring_hom {R S : Type*} [semiring R] [semiring S] (f : R → S)
  extends is_monoid_hom f :=
(map_add : ∀ x y : R, f (x + y) = f x + f y)
(map_zero : f 0 = 0)

instance id.is_ring_hom (R : Type*) [semiring R] :
  is_ring_hom (id : R → R) := sorry -- details omitted

lemma comp.is_ring_hom {R S T : Type*} (f : R → S) (g : S → T)
  [semiring R] [semiring S] [semiring T] [is_ring_hom f] [is_ring_hom g] :
  is_ring_hom (g ∘ f) := sorry -- details omitted
\end{lstlisting}

Synthesis for the \mbox{\lstinline{is_ring_hom}} class struggles with the resulting higher-order matching problems.
In particular, there is no instance for composition since matching \lstinline{is_ring_hom (?_g ∘ ?_f)} with a goal \lstinline{is_ring_hom f}
would result in setting \lstinline{?_f} to \lstinline{f} and \lstinline{?_g} to the identity function \lstinline{id}.
Thus, making \lstinline{comp.is_ring_hom} would lead to instance synthesis diverging along the path \lstinline{is_ring_hom f → is_ring_hom (id ∘ f) → is_ring_hom} \lstinline{(id ∘ id ∘ f) → ⋯}.
% Even if \lstinline{id.is_ring_hom} were not an instance, the tricky higher-order unification in synthesis lead to subtle issues.
In the formalization of Witt vectors, these issues led Commelin and Lewis to avoid classes and instead use a custom metaprogram for generating instances of their \lstinline{is_poly} predicate~\cite{Witt-vectors}.

Apart from the inability of instances on compositions to be synthesised,
under this design rewriting tactics such as the simplifier cannot easily iterate over all subterms where the \lstinline{map_mul} lemma can be applied:
since every subterm of any term can potentially unify with a function application (such as a constant function),
any subterm would cause an instance search.
Finally, the collection of morphisms could not be as easily treated as an object in its own right,
for example to put a group structure on the automorphisms of a field~\cite{mathlib}.

\subsection{Bundled morphisms}
For these reasons, \mathlib was refactored to prefer bundled morphisms:
\begin{lstlisting}
structure monoid_hom (M N : Type*) [monoid M] [monoid N] :=
(to_fun : M → N)
(map_mul : ∀ x y, to_fun (x * y) = to_fun x * to_fun y)
(map_one : to_fun 1 = 1)

structure ring_hom (R S : Type*) [semiring R] [semiring S]
  extends monoid_hom R S :=
(map_add : ∀ x y, to_fun (x + y) = to_fun x + to_fun y)
(map_zero : to_fun 0 = 0)

instance monoid_hom.has_coe_to_fun (M N : Type*) [monoid M] [monoid N] :
  has_coe_to_fun (monoid_hom M N) (λ _, M → N) :=
{ coe := monoid_hom.to_fun }

def monoid_hom.id (M : Type*) [monoid M] : monoid_hom M M :=
{ to_fun := id, .. } -- details omitted

def monoid_hom.comp {M N O : Type*} [monoid M] [monoid N] [monoid O]
  (f : monoid_hom M N) (g : monoid_hom N O) : monoid_hom M O :=
{ to_fun := g ∘ f, .. } -- details omitted
\end{lstlisting}
Lean uses the \lstinline{has_coe_to_fun} instance to parse \lstinline{(f : monoid_hom M N) x} as \lstinline{(@coe_fn _ _ (monoid_hom.has_coe_to_fun M N) f : M → N) x}.
Further examples of bundled morphisms available in \mathlib include ring homomorphisms, linear maps, monotone functions (order homomorphisms)
and the bijective versions of the above: group, ring and order isomorphisms and linear equivalences.

Bundled morphisms do not suffer from the composition, simplification and structure issues,
at the cost of all morphisms needing to be declared as such ahead of time or needing lemmas to convert between bundled and unbundled forms.
This is a drawback especially when the unbundled form has convenient notation,
such as the additive group endomorphism of a ring given by multiplying by a constant \lstinline{c}:
\begin{lstlisting}
instance mul.is_add_monoid_hom {R : Type*} [ring R] (c : R) :
  is_add_monoid_hom ((*) c) := sorry -- details omitted

def add_monoid_hom.mul_left {R : Type*} [ring R] (c : R) :
  add_monoid_hom R R := { to_fun := (*) c, ..sorry } -- details omitted
\end{lstlisting}

In addition, it is no longer possible to use \lstinline{monoid_hom} lemmas for a \mbox{\lstinline{ring_hom}:}
since \lstinline{monoid_hom} and \lstinline{ring_hom} are two different bundled types,
ring homomorphisms can be viewed as monoid homomorphisms only through (manually) inserting coercions.
Although the coercion could be supplied in some cases using unification hints, the support for unification hints in Lean 3 was not sufficient to do this in every case,
and in anticipation of Lean 4's new unification system, unification hints were entirely removed from Lean 3 and \mathlib.

Instead, to gain fully automatic simplification, all \lstinline{monoid_hom} lemmas had to be copied over to \lstinline{ring_hom} and all other structures extending \lstinline{monoid_hom}.
Thus \mathlib ended up with many copies of lemmas such as \lstinline{map_prod}:
\begin{lstlisting}
lemma monoid_hom.map_prod (g : monoid_hom M N) :
  g ( ∏ i in s, f i) =  ∏ i in s, g (f i) := sorry -- proof omitted

lemma ring_hom.map_prod (g : ring_hom R S) :
  g ( ∏ i in s, f i) =  ∏ i in s, g (f i) :=
monoid_hom.map_prod s f g.to_monoid_hom

lemma mul_equiv.map_prod ...
lemma ring_equiv.map_prod ...
lemma alg_hom.map_prod ...
lemma alg_equiv.map_prod ...
\end{lstlisting}
This duplication is further multiplied by the amount of monoid operators in \mathlib:
a corresponding version of each \lstinline{map_prod} lemma also exists for the product of a multiset and for the product of a list.
Furthermore, monoid homomorphisms preserve multiplicative inverses, powers of elements, divisibility, $n$th roots, and so on.
The end result is that the full set of lemmas grows proportionally to the number of structures extending \lstinline{monoid_hom} times the number of operations preserved by a \lstinline{monoid_hom}.

This copying happened manually and typically on an ad hoc basis, so that \mathlib contributors often encountered lemmas that were missing for their specific choice of morphism,
needing to switch contexts and add these mathematically trivial lemmas back in by hand, waiting for the dependencies to recompile before being able to continue with their proof.
To address these shortcomings, the \mathlib community on initiative of the author of this paper, switched to a third design pattern that automates the derivation of lemmas when a morphism type is extended.

\subsection{Morphism classes}
The cause of this duplication is that the pattern of ``bundled morphism'' was applied informally,
with no unifying programmatic interface.
The key insight was to follow object-oriented practice of programming to an interface rather than a concrete class, or in Lean terms:
to program to a typeclass \lstinline{monoid_hom_class} rather than a concrete type such as \lstinline{monoid_hom}.

The first step in introducing this interface was a typeclass \lstinline{fun_like} for \emph{types} of bundled (dependent) functions,
based on Eric Wieser's \lstinline{set_like} class for types of bundled subobjects.\footnote{\url{https://github.com/leanprover-community/mathlib/pull/6768}}
\begin{lstlisting}
class has_coe_to_fun (F : Type*) (α : out_param (F → Type*)) :=
(coe : Π x : F, α x)

class fun_like (F : Type*)
  (α : out_param Type*) (β : out_param (α → Type*))
  extends has_coe_to_fun F (λ _, Π a : α, β a) :=
(coe_injective' : function.injective coe)

-- A typical instance looks like:
instance monoid_hom.fun_like : fun_like (monoid_hom M N) M (λ _, N) :=
{ coe := monoid_hom.to_fun,
  coe_injective' := λ f g h, by { cases f, cases g, congr' } }
\end{lstlisting}
After defining the instance \lstinline{monoid_hom.fun_like},
instance synthesis provides function application syntax, extensionality and congruence lemmas for monoid homomorphisms.

The next step in addressing the duplication is to introduce a class for
the bundled morphism types that coerce to \lstinline{monoid_hom}:
\begin{lstlisting}
class monoid_hom_class (F : Type*) (M N : out_param Type*)
  [monoid M] [monoid N] extends fun_like F M (λ _, N) :=
(map_one : ∀ (f : F), f 1 = 1)
(map_mul : ∀ (f : F) (x y : M), f (x * y) = f x * f y)

instance : monoid_hom_class (monoid_hom M N) M N :=
sorry -- details omitted
\end{lstlisting}
Note the difference between \lstinline{is_monoid_hom f} and \lstinline{monoid_hom_class F M N}:
the former is a predicate on \emph{morphisms}, the latter is a predicate on \emph{types of morphisms}.

It is necessary to fully apply the morphism types before they can be used as a parameter to \lstinline{monoid_hom_class}:
since \lstinline{monoid_hom} and \lstinline{ring_hom} have different instance parameters,
we are not able to write both \lstinline{monoid_hom_class monoid_hom} and \lstinline{monoid_hom_class} \lstinline{ring_hom} type-correctly.
This means the class requires parameters \lstinline{M N}, which are \lstinline{out_param}s so that the lemma application \lstinline{map_one f} can leave these parameters implicit.

The types such as \lstinline{ring_hom} extending \lstinline{monoid_hom} should receive a \lstinline{monoid_hom_class} instance,
which we can do by subclassing \lstinline{monoid_hom_class} and instantiating the subclass:
\begin{lstlisting}
class ring_hom_class (F : Type*) (R S : out_param Type*)
  [semiring R] [semiring S]
  extends monoid_hom_class F R S :=
(map_zero : ∀ (f : F), f 0 = 0)
(map_add : ∀ (f : F) (x y : R), f (x + y) = f x + f y)

instance : ring_hom_class (ring_hom R S) R S := sorry -- details omitted
\end{lstlisting}

Now lemmas can be made generic by parametrizing over all the types of bundled morphisms,
reducing the multiplicative amount of lemmas to an additive amount:
each extension of \lstinline{monoid_hom} should get a \lstinline{monoid_hom_class} instance,
and each operation preserved by \lstinline{monoid_hom}s should get a lemma taking a \lstinline{monoid_hom_class} parameter.
\begin{lstlisting}
lemma map_prod {G : Type*} [monoid_hom_class G M N] (g : G) :
  g ( ∏ i in s, f i) =  ∏ i in s, g (f i) := sorry -- proof omitted
\end{lstlisting}

This design pattern has been applied in \mathlib to morphisms (implemented as subclasses of \lstinline{fun_like}) and subobjects (implemented as subclasses of \lstinline{set_like}).
The \mathlib community has welcomed the morphism class design for reducing the amounts of duplication, manual work and missing lemmas,
although not all usages have switched to the generic lemmas,
and work is still ongoing to provide a suitable generic form of standard operations such as composition and identity maps.

\section{\lstinline{nsmul}: ensuring equality of instances} \label{sec:nsmul} \label{sec:diamond}

Each \lstinline{add_comm_monoid M} structure naturally gives rise to an \lstinline{ℕ}-module structure,
where \lstinline{n • x} is defined as \lstinline{x + x + ⋯ + x}, \lstinline{n} times.
In addition, each \lstinline{semiring R} structure naturally gives rise to an \lstinline{R}-module structure on itself,
where \lstinline{x • y} is defined as \lstinline{x * y}.
These two actions are available in \mathlib as instances \lstinline{add_comm_monoid.nat_module} and \lstinline{semiring.to_module} respectively.
Note that setting \lstinline{M = R = ℕ} results in two instances for \lstinline{module ℕ ℕ}.
The existence of multiple instances of the same type does not necessarily lead to problems in Lean.
Indeed, diamond inheritance in the \mathlib algebraic hierarchy exploits this possibility.
Problems arise when the two instances are not definitionally equal, in cases such as a goal containing \lstinline{add_comm_monoid.nat_module} in which we want to apply a lemma containing \lstinline{semiring.to_module}.
As an extra complication, the two instances result in the same syntax \lstinline{n • k}, making incompatibilities hard to spot.

To resolve such issues, first we could ensure only one instance is found,
for example by replacing the other instance with a \lstinline{def} that is not considered during instance synthesis.
However, both described above are mathematically useful in their respective context,
and only cause an issue when this context overlaps, namely for the natural numbers.
Modifying the order in which instances are considered will not work, since one instance is not merely a generalization of the other:
when combining a lemma on \mbox{\lstinline{add_comm_monoid}s} with a lemma on \mbox{\lstinline{semiring}s,} both instances will still appear no matter the instance priorities.

When overlapping instances are required, the \mathlib community ensures these are definitionally equal for all possible instantiations in the overlap.
Note that Lean's implementation of diamond inheritance automatically provides definitional equality of all inheritance paths.

An advantage of the \lstinline{Prop}-valued mixin classes discussed in Section~\ref{sec:mixin} is that all instances are equal by proof irrelevance.
For example, the \mathlib community is considering replacing the data-carrying class \lstinline{fintype (α : Type*) : Type*} containing a finite enumeration of the elements of a given type,
with a proof-only class \lstinline{finite (α : Type*) : Prop} non-constructively asserting the existence of an enumeration.
Although \lstinline{fintype α} % uses quotient types to ensure that \lstinline{fintype α}
is designed to be a subsingleton for all \lstinline{α},
it is only a subsingleton up to propositional equality, meaning two different enumerations would still lead to unification issues.
On the other hand, \lstinline{Prop}-valued classes cannot be applied everywhere:
the absence of data means it is incompatible with classes that provide notation such as scalar multiplication,
and it is in general incompatible with intuitionistic logic.
The class \mbox{\lstinline|decidable_pred \{α : Type*\}|} \lstinline|(p : α → Prop) : Type*| provides a decision algorithm for \lstinline{p},
and is used in \mathlib for small numeric computations.
While we could define this to be \lstinline{Prop}-valued by setting \mbox{\lstinline{decidable_pred p := ∀ x, p x ∨ ¬ (p x)}},
that would make it useless for actually performing this decision algorithm.

To make the two \lstinline{module ℕ ℕ} instances definitionally equal,
we ensure data-carrying fields of these instances are definitionally equal, using proof irrelevance for the proof-carrying fields~\cite{mathlib-scalar-actions}.
In particular, the \lstinline{smul} field of \lstinline{add_comm_monoid.module} needs to be defined
so that instantiated for \lstinline{ℕ}, it equals multiplication on natural numbers \lstinline{nat.mul}.
While we could redefine \lstinline{nat.mul} to be recursive on the left argument to match the action of left modules,
this would violate the requirements of right modules, where multiplication by natural numbers must be right-recursive.

Instead, \mathlib adds extra data to \lstinline{add_comm_monoid}'s ancestor \lstinline{add_monoid}:
a field \lstinline{nsmul : ℕ → M → M} defines scalar multiplication by a natural number,
and two proof fields assert it (propositionally) equals the left-recursive definition: % (propositional) equalities \lstinline{nsmul 0 x = 0} and \lstinline{nsmul (n + 1) x = x + nsmul n x}:
\begin{lstlisting}
class add_monoid (M : Type*) extends add_semigroup M, add_zero_class M :=
(nsmul : ℕ → M → M)
(nsmul_zero : ∀ x, nsmul 0 x = 0)
(nsmul_succ : ∀ (n : ℕ) x, nsmul (n + 1) x = x + nsmul n x)
\end{lstlisting}
The \lstinline{nsmul} field can be set to the usual \lstinline{*} operator for \lstinline{add_comm_monoid ℕ},
and a generic implementation \lstinline|nsmul_rec {M : Type*} [has_zero M]| \lstinline|[has_add M]| \lstinline|: ℕ → M → M| is provided for instances where definitional equality is not a concern.
%As a convenience, \lstinline{nsmul_rec} is provided as a default and the equality fields are filled using the \lstinline{refl} tactic, so instantiating \lstinline{add_comm_monoid} only involves \lstinline{nsmul} when needed:
% In fact, for symmetry \lstinline{monoid} includes a field \lstinline{npow} for raising an element of the monoid to a given natural number power.

% div_inv_monoid, see \cite{class-number}

The same principle of providing a field for all definitional equalities
generalizes the principle of \emph{forgetful inheritance}~\cite{forgetful-inheritance} known also in Coq and Isabelle,
that the instance creating a superclass from a subclass can only consist of projecting away fields.
This rule is illustrated in \mathlib by the class \lstinline{metric_space} which extends \lstinline{topological_space}~\cite{perfectoid-spaces,mathlib}.

There is currently no mechanism available in \mathlib for automatically detecting or resolving issues with definitional equality of instances.
A linter~\cite{mathlib-linting} that warns for diamond issues would already be a useful improvement over the status quo of manual investigation.
Even better would be a mechanism that can canonicalize instances of propositionally subsingleton classes to ensure equality also holds definitionally.

\section{\lstinline{unique}: proof-carrying mixin} \label{sec:mixin}

The \mathlib algebraic hierarchy is \emph{semi-bundled}, meaning all operations and properties are passed in a single instance parameter.
In contrast, \mathlib also provides a large collection of mixins that can be added as separate instance parameters.
For example, \lstinline{subsingleton : Π (α : Type*), Prop} asserts the type \lstinline{α} has at most one element.
The subclass \lstinline{unique α} of \lstinline{subsingleton α} (constructively) asserts that \lstinline{α} has exactly one element.
This means \lstinline{unique α} is also a subclass of \lstinline{inhabited α}, which (constructively) specifies an element of \lstinline{α} while also allowing for more.
A theorem about trivial monoids will take these assumptions as separate parameters \lstinline{[monoid M] [subsingleton M]}:
\begin{lstlisting}
instance [monoid M] [subsingleton M] : unique (units M) :=
sorry -- proof omitted
\end{lstlisting}

In fact, \lstinline{unique α} is equivalent to the conjunction of \lstinline{subsingleton α} and \lstinline{inhabited α}.
However, the implication \lstinline!∀ {α}, subsingleton α → inhabited α → unique α! cannot be added while keeping \lstinline{subsingleton} and \lstinline{inhabited} superclasses of \lstinline{unique},
since that would result in an infinite loop \lstinline{unique → subsingleton → unique → subsingleton → ⋯} during instance synthesis.
The tabled instance synthesis procedure in Lean 4 will ensure searches are performed only once per syntactically equal subgoal, resolving this specific issue~\cite{lean-tabled-typeclasses}.
The current version of \mathlib still uses such conjunction classes even though instances cannot be automatically synthesized from conjuncts.
Preferring a single instance parameter improves performance by reducing term size, as we will discuss in Section~\ref{sec:performance}.

\section{\lstinline{fact}: interfacing between instances and non-instances} \label{sec:fact}

Suppose we want to create an instance reflecting the fact that $\Z/n\Z$ is a field if $n$ is a prime number.
This instance will take a number \lstinline{n : ℕ} and a proof showing \lstinline{n} is prime,
and return a \lstinline{field (zmod n)} instance.
Given \lstinline{n}, a proof that \lstinline{n} is prime cannot be inferred through unification,
so to make the instance synthesizable, the proof of primality must appear as an instance parameter.
Thus, we could define a class \lstinline{nat.prime n} asserting \lstinline{n : ℕ} is a prime number,
and take an instance of this class as a parameter of the \lstinline{zmod.field} instance:
\begin{lstlisting}
class nat.prime (n : ℕ) : Prop :=
(nontrivial : 2 ≤ n) (only_two_divisors : ∀ m ∣ n, m = 1 ∨ m = n)

def zmod : ℕ → Type
| 0     := ℤ
| (n+1) := fin (n+1)

instance zmod.field (n : ℕ) [nat.prime n] : field (zmod n) :=
sorry -- details omitted
\end{lstlisting}
Unfortunately, instances for \lstinline{nat.prime} do not work well in their own right:
it is impractical to check that a term \lstinline{n} is a prime number by recursion on the term structure of \lstinline{n}.
In particular, \lstinline{n} may contain free variables or be too large to reduce to a unary numeral.
Splitting between a predicate \lstinline{def nat.prime} whose proofs are passed as explicit parameters
and the same predicate as a class declaration \lstinline{class nat.prime_class} whose proofs are passed as instance parameters is not satisfying either,
due to the large amount of duplication this would entail.

Instead \mathlib provides a mechanism for ad hoc typeclass creation,
by supplying a proposition to the \lstinline{fact} class:
\begin{lstlisting}
def nat.prime (n : ℕ) : Prop := 2 ≤ n ∧ (∀ m ∣ n, m = 1 ∨ m = n)

class fact (p : Prop) : Prop := (out : p)

instance zmod.field (n : ℕ) [fact (nat.prime n)] : field (zmod n) :=
sorry -- details omitted
\end{lstlisting}

In a similar way, the \lstinline{fact} class is used for the assumption \lstinline{x < y} when showing that the interval $[x, y] \subset \R$ is a manifold with boundary,
to provide the assumption that a polynomial $f$ splits in a field $K$ when defining the natural inclusion of the splitting field of $f$ into $K$,
and to provide non-negativity or positivity assumptions in various contexts.

Along with the ad hoc class pattern provided by \lstinline{fact},
there is an ad hoc instance pattern provided by the tactic \lstinline{letI}.
Instance synthesis considers declarations marked as \lstinline{instance} and parameters to the current declaration, caching these before elaborating the type and body of the declaration.
The \lstinline{letI} tactic inserts new instances into this cache,
providing this instance in the current proof context.

In addition, \lstinline{letI} can resolve the dangerous instance issue of Section~\ref{sec:dangerous-instance} in some cases:
in a proof context where the ring of scalars \lstinline{R} remains fixed, we can use \lstinline{letI} to safely make \lstinline{module.add_comm_monoid_to_add_comm_group} an instance within this context.

\section{Performance and bundling} \label{sec:performance}

The pervasive use of typeclasses in \mathlib means instance synthesis accounts for 10 to 25 percent of the build time of a typical \mathlib file. % TODO: get a beefier computer so we can run the profiler on the entirety of mathlib rather than a random selection of mathlib files
Beyond the time taken for synthesis, the use of typeclasses has performance impacts on the entirety of the compilation process.
For example, typeclasses tend to produce larger terms compared to those generated by canonical structure:
Another factor complicating direct comparison with other mechanisms is the trade-off between upfront and repeating costs.
For instance, although activating a locale in Isabelle is rather costly since it requires processing all declarations in the locale's dependency graph~\cite{ballarin-locales},
declarations can be structured such that this cost only needs to be paid once for a group of declarations;
instance synthesis is performed for each instance parameter in each term.
Performance of typeclasses in Lean remains acceptable, though, thanks to instance caching, Lean's efficient synthesis implementation in C++ and \mathlib's design patterns.

First of all, the \mathlib algebraic hierarchy avoids unbundled subclasses that express superclass constraints through instance parameters,
since these lead to exponential blowup of term sizes.
An example of exponential blowup is discussed in detail in Ralf Jung's blog post~\cite{unbundled-blowup}.
This example concerns product type instances of an unbundled class such as the following modification to the \lstinline{comm_monoid} class:
\begin{lstlisting}
class semigroup (G : Type*) [has_mul G] := ...
class mul_one_class (M : Type*) [has_one M] [has_mul M] := ...
class comm_semigroup (G : Type*) [semigroup G] := ...
class monoid (M : Type*) [semigroup M] [mul_one_class M].
class comm_monoid (M : Type*) [monoid M] [comm_semigroup M].
\end{lstlisting}

Providing an instance for the natural numbers is straightforward, although it now involves instantiating each step in the hierarchy separately:
\begin{lstlisting}
instance : semigroup ℕ := sorry -- details omitted
instance : mul_one_class ℕ := sorry -- details omitted
instance : comm_semigroup ℕ := sorry -- details omitted
instance : monoid ℕ := sorry -- details omitted
instance : comm_monoid ℕ := sorry -- details omitted
\end{lstlisting}

When we want to instantiate the commutative monoid structure on the product of two commutative monoids, we see that the length of types starts to grow noticeably:
\begin{lstlisting}
instance prod.has_mul [has_mul G] [has_mul H] : has_mul (G × H) :=
{ mul := λ a b, (a.1 * b.1, a.2 * b.2) }
instance prod.semigroup [has_mul G] [has_mul H]
  [semigroup G] [semigroup H] : semigroup (G × H) :=
sorry -- details omitted
...
instance prod.comm_monoid
  [has_one M] [has_one N] [has_mul M] [has_mul N]
  [semigroup M] [semigroup N] [mul_one_class M] [mul_one_class N]
  [monoid M] [monoid N] [comm_semigroup M] [comm_semigroup N]
  [comm_monoid M] [comm_monoid N] :
  comm_monoid (M × N) :=
sorry -- details omitted
\end{lstlisting}

The linear growth in the types translates to an exponential growth in the term size of concrete instances,
since each instance parameter implicit in \lstinline{comm_monoid (ℕ × ⋯ × ℕ)} is filled with a term that has itself the same number of instance arguments.

The performance issue of unbundled classes is well known in the Coq community since Coq has a similar implementation of classes to Lean,
and the first Coq library using classes for its algebraic hierarchy suffered from slowdowns due to this design issue~\cite{Coq-typeclass-hierarchy}.
The \emph{packed classes} design pattern used for performant canonical structures~\cite{Coq-Packed-classes} translates to bundled subclassing:
in packed classes, the substructure relation is expressed by declaring the superstructure as an instance, instead of a parameter to the record.
Similarly, \mathlib prefers bundled classes, expressing the subclass relation through incorporating the superclass as an instance, instead of a parameter on the class's type.

In addition, the deprecation of the \lstinline{old_structure_cmd} option results in improved performance for unification of instances in the presence of large inheritance chains.
Since equality of structures is determined field-wise, incorporating a parent as a field means instances deriving from the same parent instances can be immediately verified to be equal,
compared to the \lstinline{old_structure_cmd} situation where this comparison has to be performed on the union of all fields of all ancestor structures, unfolding all intermediate projections.

An important source of slowdowns is failing instance searches since the entire search space has to be exhaustively checked before failing.
If a user omits a hypothesis by mistake, the error message should not be a timeout but instead point out that the instance was not found.
Thus, all instances in \mathlib are checked by a \lstinline{fails_quickly} linter,
that checks that within an acceptable time (configurable in the linter) synthesis fails to synthesize a given instance
when arguments are missing.
The \lstinline{fails_quickly} linter can also detect timeouts caused by looping or diverging synthesis,
for example the loop \lstinline{nonempty → has_bot → nonempty} in the following code:
\begin{lstlisting}
-- `has_bot.bot` is notation for the minimum element of `α`
class has_bot (α : Type*) := (bot : α)

instance has_bot_nonempty (α : Type*) [has_bot α] : nonempty α :=
⟨has_bot.bot⟩

-- The natural numbers are well-ordered.
instance nat.subtype.has_bot (s : set ℕ) [decidable_pred (∈ s)]
  [h : nonempty s] : has_bot s := sorry -- proof omitted
\end{lstlisting}

On the other hand, sometimes failing to synthesize instances should not cause an error,
especially in tactics which can handle synthesis failures by switching to a less powerful procedure.
In particular, the simplification procedure used instance synthesis to determine whether the type of a subterm had a \lstinline{subsingleton} instance,
which would allow more powerful rewriting in the presence of dependencies between subterms.
This meant the simplification tactic searched for \lstinline{subsingleton} instances for each subterm that another subterm depended on,
in the worst case spending more than half its time on failing instance synthesis calls.
Modifying the simplification procedure to instead rely on user-supplied congruence lemmas\footnote{\url{https://github.com/leanprover-community/lean/pull/665}}
resulted in a speedup of approximately 15 percent over the entirety of \mathlib.

The depth-first nature of instance synthesis means it is advantageous to try instances that succeed or fail fast
before ones that require traversing a full tree before determining their success.
The \lstinline{priority} attribute of instances controls the order in which instances of the same class are considered:
higher priorities are tried before lower priorities.
The rule of thumb used in \mathlib states to assign a low priority to \emph{blanket instances}: those where all explicit parameters to the class are free variables.
In particular, all subclass instances are automatically assigned a lower priority.
A subtler case involved unification of quotient types:
the instance \lstinline{con.quotient.decidable_eq} states equality is decidable on the quotient of a type \lstinline{M} by any decidable multiplicative congruence relation of type \lstinline{con M}.
\begin{lstlisting}
instance con.quotient.decidable_eq {M : Type*} [has_mul M] (c : con M)
  [∀ (a b : M), decidable (c a b)] : decidable_eq (quotient c) :=
sorry -- proof omitted
\end{lstlisting}
A value for the instance parameter \lstinline{[has_mul M]} has to be synthesized when \lstinline{con.quotient.decidable_eq} is considered as candidate instance,
meaning this instance will cause a search through all instances of all subclasses of \lstinline{has_mul}.
Since \mathlib makes extensive use of other quotient types such as \lstinline{multiset α}, the quotient of \lstinline{list α} modulo permutations,
specialized instances such as \lstinline{multiset.decidable_eq : decidable_eq α → decidable_eq (multiset α)} are assigned higher priority than the expensive instance \lstinline{con.quotient.decidable_eq}.

While the above design guidelines have allowed the growth of \mathlib's class hierarchy,
the fact they often need to be verified and applied manually shows that performance is a key consideration in the further growth of the library.

\section{Conclusion}

The pervasive use of class-based patterns throughout \mathlib demonstrates that the instance parameter mechanism scales to a large interconnected library of mathematical structures.
In addition to algebraic, order and topological hierarchies, classes have proved useful for representing morphisms and subobjects.
Still, the choice between bundled and unbundled subclassing and the duplication in the additive and multiplicative hierarchy are drawbacks of the use of typeclasses that do not appear in canonical structures or locales.
For newcomers, notation for classes such as vector spaces is surprising.
Even worse, dangerous instances, definitional equality and divergence are regular sources of errors that require a good understanding of the synthesis mechanism to resolve,
and keeping the whole system performant is a permanent source of concern.

Lean 4 brings tabled instance synthesis to improve performance,
and linters are able to report both dangerous instances and divergence.
Future work should build on these improvements by providing a user friendly way of resolving definitional equality issues,
be it a linter or a way to incorporate equality into the synthesis mechanism.
In addition, macros that transform binder lists can address some of the unfamiliar notations.
Another avenue to address the drawbacks of typeclasses is the integration of classes with another hierarchy mechanism, such as the combination of classes and locales in Isabelle.

\bibliography{lean}

\end{document}